\documentstyle[preprint,prl,aps]{revtex}

\begin{document}
\title { Nonlinear differential equations based on nonextensive 
Tsallis entropy and physical applications}
\author { R. S. Mendes and I. T. Pedron}
\address { Departamento de F\'{\i}sica, 
Universidade Estadual de Maring\'a, \\
Av. Colombo 5790, 87020-900 Maring\'a-PR, Brazil}
\date{\today }
\maketitle
\draft
\begin{abstract}
A family of nonlinear ordinary differential equations  with arbitrary
order is obtained by using nonextensive concepts related
to the Tsallis entropy. Applications of these equations are
given here. In particular, a connection between 
Tsallis entropy and the one-dimensional correlated anomalous diffusion 
equation is established. 
It is also developed explicitly a WKB-like 
method for second order equations and it is applied 
to solve approximately  a class of equations that contains as 
a special case the Thomas-Fermi equation for an atom.
It is expected that the present ideas can be useful
in the discussion of other nonlinear contexts.  
\end{abstract}


\date{\today}


\section{Introduction}
As it is well known, the equation
\begin{eqnarray}
\frac{dp}{dx} + \lambda\; p=0\; ,
\label{a1}
\end{eqnarray}
whose general solution  is 
\begin{eqnarray}
p(x)=p_{o}\exp(-\lambda x)\; ,
\label{po}
\end{eqnarray} 
has many applications. In general,  Eq. (\ref{a1}) 
models  systems related to an {\it extensive} context (in the thermodynamical sense).
Typical examples are the systems based on independent events, 
for instance, the radioactive decay of noninteracting nucleus.
In this context, it is interesting to observe 
that the solution of Eq. (\ref{a1}) has an entropic interpretation.
In fact,  if the usual entropy,
\begin{eqnarray}
S = - \int_a^{b} p(x)\; \ln p(x) \;{\mbox d}x \; ,
\label{a2}
\end{eqnarray}
is maximized subject to the constraints
\begin{eqnarray}
\alpha= \int_a^{b}  p(x)\; {\mbox d}x 
\label{a3}
\end{eqnarray}
and
\begin{eqnarray}
\beta=\int_a^{b} x \;p(x)\; {\mbox d}x \; ,
\label{a4}
\end{eqnarray}
the solution of Eq. (\ref{a1}) for $a\leq \;x\;\leq\;b$ is obtained again. 
The parameters $\alpha$ and $\beta$ are adjusted 
in order to obtain $p_o$ and $\lambda$.
If $\alpha=1$,  $p(x)$ can be interpreted as a probability.
However, this choice for $\alpha$  is not necessary because 
all the conclusions presented here are independent of the $\alpha$ value. 

Now, it is possible to ask how  Eq. (\ref{a1}) 
must be generalized in order to describe 
dependent events ({\it nonextensive} context).
As a guide to answer this question it will 
be considered in this work a 
generalization of the entropy (\ref{a2})
employed  by Tsallis\cite{ts88} (see also Ref. \cite{curado})  
in a nonextensive statistical mechanics.
The above choice is motivated by the fact that this 
generalized entropy (Tsallis entropy) has been
applied successfuly in the discussion of many
situations where the nonextensivity plays an important role,
for instance, L\'evy superdiffusion\cite{zanette}  
and correlated anomalous diffusion\cite{plastino,tsallis,borland},
turbulence in two-dimensional pure electron plasma\cite{boghosian},
dynamic linear response theory\cite{rajagopal1} and Green functions\cite{mendes},
perturbation and variation methods for calculation of thermodynamic
quantities\cite{lenzi},
low-dimensional dissipative systems\cite{lyra},
simulated annealing and optimization techniques\cite{moret} and 
connection with quantum uncertainty\cite{rajagopal2}.
In other words, the main purpose of this work is to obtain generalizations of
Eq. (\ref{a1}) based on nonextensive Tsallis entropy,
giving physical applications.
More specifically, it is obtained a generalization of Eq. (\ref{a1})
by using  Tsallis entropy (Section {\bf II}), and  applications to 
the motion of a particle in a fluid medium and chemical kinetics 
are presented. 
A further generalization for higher order nonlinear differential equations 
is given (Section {\bf III}) and it is applied to discuss a 
new connection  between the
correlated anomalous diffusion equation  
and  the Tsallis entropy.
Moreover, it is introduced a WKB-like approximation in order to study
a family of nonlinear differential equations based on 
Tsallis entropy (Section {\bf IV}), and finally this approximated method
is employed to the case of the Thomas-Fermi equation.


\section{First order nonlinear differential equation}
The Tsallis entropy can be written as
\begin{eqnarray}
S_q = - \int^{b}_{a} \frac{p(x) \left (  
1- p(x)^{q-1} \right )}{1-q} \; {\mbox d}x \; .
\label{a5}
\end{eqnarray}
Furthermore, the constraint (\ref{a4}) is currently  
substituted\cite{ts88,curado} by
\begin{eqnarray}
\beta = \int_{a}^{b} x\; p(x)^q \;{\mbox d} x  
\label{a6}
\end{eqnarray}
and the constraint (\ref{a3}) remains unchanged. 
When the entropy (\ref{a5}) is maximized subject to the 
constraints (\ref{a3}) and (\ref{a6}), $p(x)$ can be written as 
\begin{eqnarray}
p(x) = p_o \left[1-(1-q)\;p_o^{q-1}\; 
\lambda x \right]^{1 / (1-q)} \; .
\label{a7}
\end{eqnarray}
This distribution  generalizes the exponential one (see Eq.  (\ref{po})).
In the above expressions the parameter $q \; \in \; {\cal R}$ 
characterizes the degree of  nonextensivity, in particular  
the entropy (\ref{a2}) and the constraint (\ref{a3})
are obtained as limiting case when $q \rightarrow 1$, 
recovering the extensive case. 
A remarkable fact is that Eq. (\ref{a7}) can be applied 
directly to several contexts successfuly.
In addition to other works already cited, the study of  
solar neutrinos\cite{kaniadakis} and Zipf law\cite{denisov}
are relevant examples of such applications.

By direct inspection it is  verified that the above 
generalization of the exponential satisfies the 
nonlinear equation
\begin{eqnarray}
\frac{dp}{dx}+\lambda \; p^q=0
\label{a8}
\end{eqnarray}
subject to the initial condition $p(0) = p_o$.
This equation is the generalization of Eq. (\ref{a1})
based on the nonextensive Tsallis entropy.
When the notation  $\lambda_{eff}= \lambda p^{q-1}$ is employed,
 Eq. (\ref{a8}) can be interpreted as a decay with  memory,
thus $q$ is in some sense  related with the memory of the system.


\subsection{Applications}
Before  generalizing Eq. (\ref{a8}) to higher orders 
it is  illustrative to present applications for it.
One of them describes the mean motion of a particle in a fluid medium 
without external force. In this case, the motion equation can be written as  
\begin{eqnarray}
m{\mbox d}v / {\mbox d}t = - b\, v^{q} , 
\label {delta1}
\end{eqnarray}
where $v$ and $b$
are respectively the velocity and the friction coefficient of the particle 
 relative to the medium and $m$ is the particle mass. Here,  the parameter $q$
is related to the turbulent flow (nonextensive
behavior) and  for $q=1$  (slow motion)  the extensive behavior is recovered.
Another example comes from chemical kinetics.
In this case, the concentration $C_A$ of a given species $A$ 
obeys the empirical equation\cite {quimica}
\begin{eqnarray}
 {{\mbox d}C_A}/{{\mbox d}t}=K \, C_{A}^{\alpha}\, 
C_{B}^{\beta}\, C_{C}^{\gamma} \cdot\cdot\cdot ,
\label{delta2}
\end{eqnarray}
where $K$ is the reaction constant and 
$\alpha$, $\beta$, $\gamma$, $\cdot \cdot \cdot$, 
refer to the concentration of chemical species $A$, $B$, $C$, $\cdot \cdot \cdot$,  
present in the reaction. In this expression, $\alpha$, $\beta$, $\gamma$, 
$\cdot\cdot\cdot$,  are respectively the order of the reaction with respect to  
$A$, $B$, $C$, $\cdot\cdot\cdot$, 
and the sum $\alpha+\beta+\gamma+ \cdot \cdot \cdot$ is the
overall order of the reaction.   
In some cases the concentrations $C_B$, $C_C$, ... can be considered constant, 
thus the above equation reduces to the form of Eq. (\ref{a8}) with $q=\alpha$. 
In this way, the parameter $q$ becomes the order of the reaction for the species $A$.


\section{Family of nonlinear differential equations}
The natural generalization of Eq. (\ref{a1}) for higher order differential
equations, in the sense that $p=p_{o}\exp(-\lambda x)$ is a particular solution, 
is the family of linear differential equations of 
arbitrary order  with constant coefficients. 
The $N$-order element of this family  is
\begin{eqnarray}
\sum_{n=0}^{N}a_n \frac{{\mbox d}^n p}{{\mbox d} x^n} = 0 
\label{ta}
\end{eqnarray}
and the corresponding algebraic equation for $\lambda$ is
\begin{eqnarray}
\sum_{n=0}^N a_n(-\lambda)^n = 0 \;\; .  
\label{tb}
\end{eqnarray}
In view of the previous remarks, the generalization of these ideas 
in the Tsallis context
is based on the replacement of the particular solution $p(x)=p_{o}\exp(-\lambda x)$ by
 $p(x) = p_o[1 - (1-q) p_o^{q-1}\lambda x]^{1/(1-q)}$.
Thus, 
\begin{eqnarray}
\sum_{n=0}^{N}a_n\frac{\mbox d^n}{ {\mbox d}x^n}p^{(N-n)(q-1)+1} = 0
\label{tc}
\end{eqnarray}
represents the  nonlinear ordinary  
differential equation of  constant coefficients that generalizes 
Eq. (\ref{ta}) and the corresponding generalization of Eq.(\ref{tb}) is
\begin{eqnarray}
a_N(-\lambda)^{N} \prod_{j=0}^{N-2}[(j+1)q-j] + 
\sum_{n=1}^{N-1} a_n(-\lambda)^n  \prod_{j=N-n-1}^{N-2}[(j+1)q-j] + a_0 = 0
\;\;.
\label{td}
\end{eqnarray}
This equation is applicable for $N\geq 2$ and when $N=1$  
Eq.(\ref{td}) must be replaced by $a_1(-\lambda) + a_0=0$.
Without loss of generality the $N$-th term of Eq. (\ref{tc}), 
$\mbox d^N p/\mbox dx^N$,
was considered linear in $p(x)$.
As expected, Eqs. (\ref{tc}) and (\ref{td}) reduce respectively to
Eqs. (\ref{ta}) and (\ref{tb}) in the limit $q\rightarrow 1$. Furthermore,
 it is important to remark that a superposition of particular solutions
of Eq. (\ref{tc}) is not another solution of this equation, in contrast with  the
case of Eq. (\ref{ta}). 
For instance, the generalization of circular and hyperbolic functions
based on the superposition of Eq. (\ref{a7}) \cite{borges} is not a solution
of Eq. (\ref{tc}).
Thus, in the following applications only particular solutions 
previously described will be considered.


\subsection{Application to correlated anomalous diffusion}
It was recently shown that the one-dimensional correlated anomalous diffusion 
equation (generalized nonlinear Fokker-Planck equation without external force),
\begin{eqnarray}
\frac{\partial \phi}{\partial t }=\frac{\partial^2\phi^{\nu}}{\partial x^2} \;\;\;
(\nu\in \cal R)\; ,
\label{te}
\end{eqnarray}
has a solution  related to the Tsallis entropy \cite{plastino,tsallis}.

This solution is a generalization of the Gaussian one, reducing to it 
in the limit $q\rightarrow 1$.
On the other hand, by using the generalization developed above  
for linear ordinary differential  equations, 
a new relation between Tsallis  entropy
and correlated anomalous diffusion equation can be obtained.

The new  connection is based on the substitution of 
\begin{eqnarray}
\phi(x,t) = T(t)\, X(x) 
\label{delta4}
\end{eqnarray}
into Eq. (\ref{te}).
This  separation of variables leads to 
${ {\mbox d} T}/{ {\mbox d} t} = -\sigma \;T^{\nu}$
and
${ {\mbox d}^2 Y}/{ {\mbox d}x^2} = -\sigma \;Y^{1/\nu}$,   
where $\sigma$ is a separation variable constant  and $Y = X^{\nu}$.
Since these equations  are members of the family 
ruled by  Eq. (\ref{tc}), a new connection
between the one-dimensional correlated anomalous diffusion equation and the 
Tsallis entropy is established. 
Furthermore, as a consequence of this general procedure,
 a set of particular solutions of the 
anomalous correlated diffusion equations is obtained, 
namely,
\begin{eqnarray}
\phi (x,t)=T_{0} \left[ 1-(1-\nu)\;T_{0}^{\nu-1}\; \sigma \;t  \right]^{1/(1-\nu )}\; X_{0}
\left[ 1+\left(\frac{1-\nu}{2\nu}\right) X_{0}^{(1- \nu )/2} \lambda\;  x  \right]^{2/(\nu -1)} \;\;,
\label{delta5}
\end{eqnarray}
where $T_0$ and $X_{0}$ are constants and $\lambda$ is the solution of the equation
$(1+\nu) \lambda^{2}+2 \nu \sigma=0$.
Note that the above procedure leads to complex solutions for $\sigma >0$ 
and it is a natural extension of the method of separation of variables
applied to the usual diffusion equation ($\nu =1$). 

The previous method can be easily applied to 
 the generalized diffusion equation recently proposed  
by Tsallis and Bukman\cite{tsallis},
\begin{eqnarray}
\frac {\partial {\phi}^{\mu}} {\partial t}=\frac {\partial ^2 {\phi}^{\nu}}{\partial x^2}
\;\;\;\; (\mu, \nu \in {\cal{R}})\; .
\label{delta7}
\end{eqnarray}
In fact, it is sufficient to replace $\phi ^{\mu}$ by $\psi$ and $\nu$ by $\nu / \mu$. 
In general, the connection between the porous media diffusion 
equation and Tsallis entropy, based on separation of variables and the family
of nonlinear partial differential equation (Eq. (\ref{tc})), 
can be easily extended for other nonlinear
partial differential equations.


\section{ Nonlinear differential equations and WKB-like approximation}
Eq. (\ref{a8}) can be generalized further if we 
allow  $\lambda$ to  become a function of $x$. 
In this case, the solution of Eq. (\ref{a8}) with 
$p(0)=p_o$ becomes
\begin{eqnarray}
\label{a9}
p(x)= p_o \left[ 1-(1-q)p_o^{q-1} 
\int_0^x \lambda(z) {\mbox d}z \right ]^{1/(1-q)} \; .
\end{eqnarray}
In a similar way, we can allow that the constants 
$a_n$ in Eq.(\ref{tc}) become  functions of $x$. 
This procedure leads to 
\begin{eqnarray}
\sum_{n=0}^{N}a_n(x)\frac{ {\mbox d}^n}{ {\mbox d}x^n}
p^{(N-n)(q-1)+1}=0\; .
\label{ti}
\end{eqnarray}
As in the linear case ($q=1$) there are no general solutions for these equations. 
Consequently, it is natural to perform approximated analyses to obtain some information
about the solutions of Eq.(\ref{ti}). 
In the following discussions, among other possibilities,  
a generalization of WKB method for $q\neq 1$ 
is developed explicitly for the $N=2$ case.



As it is well known,  in the WKB method\cite{wkb} 
approximate solutions of equation
${{\mbox d}^2p}/{{\mbox d}x^2}= f(x)p$
can be obtained when $f(x)$ is a slowly varying function.
In this case, it is employed  an auxiliary function $g(x)$ 
defined by the relation  
$p(x)=\exp{(g(x))}$.
>From the previous developments, a natural generalization 
of these equations are respectively 
\begin{eqnarray}
\frac{{\mbox d}^2p}{{\mbox d}x^2}= f(x)p^{2q-1}
\label{b11}
\end{eqnarray}
and
\begin{eqnarray}
p(x)=[1+(1-q)g(x)]^{1/(1-q)} \; .
\label{a13}
\end{eqnarray}

In terms of $g(x)$, Eq. (\ref{b11}) becomes
\begin{eqnarray}
[1+(1-q) g]\frac{{\mbox d}^2 g}{{\mbox d}x^2} + q 
\left( \frac{{\mbox d} g}{{\mbox d}x}\right)^2 - f = 0 \; .
\label{a14}
\end{eqnarray}
Following again the usual WKB approach, the term with ${{\mbox d}^2 g}/{{\mbox d}x^2}$ 
is neglected in the first approximation, hence
\begin{eqnarray}
g(x)= \pm q^{-1/2} \int f(x)^{1/2} {\mbox d} x \; .
\label{a16}
\end{eqnarray}
Consistently, the validity condition of this approximation is
\begin{eqnarray}
\left| \frac {{\mbox d}^2 g}{{\mbox d}x^2}\right| \approx 
\left| \frac {{{\mbox d} f}/{{\mbox d}x}}{2q^{1/2}f^{1/2}} \right|
\ll \left| \frac {f}{1+(1-q)g} \right| \;.
\label{a17}
\end{eqnarray}
Since Eq. (\ref{b11}) is nonlinear for $q\neq 1$, the 
superposition  of  solutions is not a solution,
therefore the present development is indicated for situations where some
particular solutions can be considered as good approximations.
Notice also that the solutions (\ref{a16}) can be improved
through iterations  (replacing successively improved Eq. (\ref{a16}) into Eq. ({\ref{a14})). 


\subsection{Application to Thomas-Fermi equation}
To exemplify the previous development, we consider the
Thomas-Fermi equation for a free
atom\cite{landau}
\begin{eqnarray}
\frac{{\mbox d}^{2}y}{{\mbox d}x^2}= x^{-1/2} \; y^{3/2} \; .
\label{a19}
\end{eqnarray}
In the free neutral atom case  the boundary conditions 
are $y(0)=1$ and $y(\infty)=0$.
By comparing Eq. (\ref{a19}) with Eq. (\ref{b11}) it is 
verified that $f(x)=x^{-1/2}$ and $q=5/4$.
Choosing  the particular solution adjustable 
to these  conditions it becomes
\begin{eqnarray}
y(x)=\left(1+\frac{2}{3 \sqrt{5}}\; x^{3/4}\right)^{-4} \; .
\label{a20}
\end{eqnarray}
In this example, the validity condition (\ref{a17}) becomes
\begin{eqnarray}
1+\frac {3\sqrt{5}} {2}\; x^{-3/4} \ll 15 \; .
\label{a21}
\end{eqnarray}
This condition indicates that the approximation is better
for larger $x$.
Furthermore, Eq. (\ref{a20}) is in satisfactory 
agreement with a  numerical calculation 
(see, for instance, Ref. \cite{landau}).
In a general context, when $f(x)$ is a smooth function, the 
corresponding approximate solution becomes more accurate.
In particular, when $f(x)$ is a constant this solution becomes exact.

\section{Conclusions}
Summing up, the nonextensive concepts based on the Tsallis entropy
were employed to obtain a family of nonlinear 
ordinary differential equations. 
The first order equation of this family is a nonextensive 
generalization of the exponential decay equation.
Moreover, by using a separation of variables procedure 
and the above family of equations, a connection between
the correlated anomalous diffusion 
equation and the Tsallis entropy is obtained.
In addition to this, for second order equations we presented a WKB-like 
approach to obtain approximated solutions.
This procedure was used in the context of  the 
Thomas-Fermi equation for  a free neutral atom, and it was shown that
the well known solution precisely correspond to $q=5/4$.
In general, the developments introduced in this work indicate that 
many nonlinear effects are closely related with nonextensive 
concepts in the Tsallis framework. 


\acknowledgements
I. T. Pedron thanks CAPES (Brazilian Agency) for financial support.



\end{document}